\documentclass[eng]{ajceam-class}
\title{Morphological Characterization of ABS and PC-ABS Surfaces for Automotive Industry}

\shorttitle{Author instructions for the AJCEAM}
       
\author[1]{Gianluca Deninno}
\author[1]{Ettore Vittone}

\affil[1]{University of Turin, Department of physics, Via Pietro Giuria, 1, 10125 Turin, Italy}

\author[2]{Nello Li Pira}
\author[2]{Luca Belforte}
\author[2]{Fabio Scaffidi Muta}
\author[2]{Renzo Costa}

\affil[2]{Centro Ricerche FIAT SCpA, Strada Torino 50, 10043 Orbassano, Turin – Italy}
\firstauthor{Deninno G.}

\email{gianluca.deninno@edu.unito.it} 
\receptiondate{12/10/2021}
\acceptancedate{26/01/2022}
\usepackage{cite}

\abstract{In the automotive industry, the measurement, control and reproducibility of certain morphological characteristics of surfaces are important to characterize, qualify, certify and optimize different physical properties, such as the degree of wear, the degree of adhesion by adhesives, glues or paints and optical properties. In this regard, this work is intended demonstrate how it is possible to characterize different surfaces in ABS (Acrylonitrile Butadiene Styrene) and PC-ABS (PolyCarbonate/Acrylonitrile Butadiene Styrene) through a set of morphological parameters defined according to the ISO 25178 standard. These surfaces have been subjected to different etching treatments performed with different chemicals, which can radically modify the morphology of the surface, making it more or less suitable according to the industrial purposes sought. Furthermore, in this work an analytical relationship between the hybrid parameters Sdq and Sdr defined in the ISO-25718 standard is also proposed, valid for many surfaces of general use.
This study is based on the use of a confocal optical microscope and a SEM (Scanning Electron Microscope), which allow surface morphology to be studied at the micro/nano scale and the measurements are analyzed with special imaging and three-dimensional scanning software that allows the precise measurement of morphological parameters of the standard ISO-25178.}


\begin{document}

% Include title, authors, abstract, etc.
\maketitle
%\thispagestyle{fancy}

% Main body of manuscript
\section{Introduction}
\firstword{I}{n} recent years, the development of a method to study textured surfaces has been essential as it can explain a large number of physical properties often required for industrial purposes, such as anti-corrosive, hydrophobic, hydrophilic and anti-freeze properties\cite{lv2015stability, chen2018bioinspired,wang2015superhydrophobic, patzelt2016anti}. A precise surface characterization is increasingly required in many industrial fields, from aerospace to photovoltaic industry, making a morphological characterization method indispensable\cite{macaulay2016assessment, yu2010automatic}. Even the optical properties of a surface can strongly depend on the morphology: the perceived optical effect, which is often described in the literature through the BRDF (Bidirectional Reflectance Distribution Function), is influenced by the distribution of the normal unit vectors of the surface itself, as well as by its morphology\cite{torrance1967theory, oren1995generalization, walter2007microfacet, heitz2016multiple, heitz2014importance, heitz2014understanding, smith1967geometrical, meneveaux2017rendering, simonot2009photometric}.\\
The ISO 25178  standard series of the Geometrical Product Specifications (GPS) system proposes a structured method for the characterization of the surface texture. In this standard, many morphological parameters are defined. These parameters describe specific characteristics of a surface and their measurements are based on contact and non contact techniques for the morphological characterization of the surfaces. This standard also defines several morphological filters, defined in the wavelength space, which through the calculation of the Fourier transform allows, for example, the mitigation of effects of errors arising during the measurement execution. \\ In this work several results about a restricted number of these parameters will be shown, studied and described: these parameters are able to provide a precise description of the surface morphology and through them it will be shown how it is possible to characterize a surface for many practical industrial aspects. \\The main purpose of this work is to show how it is possible to qualified chemical etched ABS and PC-ABS surfaces  through the ISO 25178 standard. In this paper, surfaces in ABS and PC-ABS material etched through chemical baths based on hexavalent chromium are studied. One of the outcome of this paper is to show the effects of different chemicals in the morphological properties of these surfaces. In particular, the methodology presented in this paper is demonstrated to be suitable for studying the morphological effects of the etching procedure carried out by hexavalent chromium and other less harmful chemicals. 

\section{Standard definitions}
\label{section:1}
\firstword{T}{he}% Capital letter in first word
parameters of the ISO 25178 standard can be divided into height parameters, volume parameters, hybrid parameters, and functional parameters\cite{isoparametri25}. In this section, the definitions of the morphological parameters are shown and additional definitions are provided. \\ The studied surface is immersed in a Cartesian reference system and it is parameterized through the graph of a two-dimensional function $f(x,y):\mathbb{R}^2\to\mathbb{R}$ of class $C^1$: in this way, it is possible to obtain the coordinates ($x, y, z$) of all the measured points through the Cartesian surface
\begin{equation}
\label{eqn:1}
    \overrightarrow{S_{\Sigma}}=(x,y,f(x,y))
\end{equation}where the \textit{x}-axis and \textit{y}-axis represent respectively the longitudinal and transverse axes of the measured surface. The mean $\overline{f}(x,y)$ of the function $f(x,y)$ appearing in relation (\ref{eqn:1}) is equal to 0, i.e. $\overline{f}(x,y)=0$.
\subsection*{Heights parameters}
The definitions of the height parameters are given in the following sections.
\subsubsection{Sq}
The \textit{Sq} parameter is the root mean square of heights of the measured points and it is defined as
\begin{equation}
    Sq = \sqrt{\frac{1}{A}\int\int_Af^2(x,y)dxdy}
\end{equation}where $A$ is the sampling area.
\subsubsection{Sz}
The \textit{Sz} parameter is defined as
\begin{equation}
    Sz = f_{MAX}(x,y)-f_{min}(x,y)
\end{equation}where $f_{MAX}(x,y)$ and $f_{min}(x,y)$ are respectively the maximum and the minimum value of the function $f(x,y)$.
\subsubsection{Ssk}
The \textit{Ssk} is the skewness parameter and it is the ratio of the mean of the height values cubed and the cube of \textit{Sq} parameter
\begin{equation}
    Ssk=\frac{1}{Sq^3}\frac{1}{A}\int\int_Af^3(x,y)dxdy
\end{equation}The values of the \textit{Ssk} can be positive, negative or zero, and is unit-less since it is normalised by \textit{Sq}.
\subsubsection{Sku}
The \textit{Sku} parameter is defined as the ratio of the mean of the fourth power of the height values and the fourth power of \textit{Sq}
\begin{equation}
    Sku = \frac{1}{Sq^4}\frac{1}{A}\int\int_Af^4(x,y)dxdy
\end{equation}As the skewness parameter, the \textit{Sku} parameter is unit-less and the values it can assume are strictly positive.
\subsection*{Hybrid parameters}
\subsubsection{Sdq}
Through the definition of the Cartesian surface shown in the equation (\ref{eqn:1}), it is possible to calculate the gradient of the surface for any points of the surface. The \textit{Sdq} parameter is defined as the root mean square of gradient and it is calculated as
\begin{equation}
    Sdq=\sqrt{\frac{1}{A}\int\int_A\left(f_x^2+f_y^2\right)dxdy}
\end{equation}where the used quantities $f_x$ and $f_y$ are respectively the partial derivative of $f(x,y)$ calculated along the \textit{x}-axis and \textit{y}-axis.
\subsubsection{Sdr}
The \textit{Sdr} parameter is the developed interfacial area ratio and it is defined as
\begin{equation}
    Sdr=\frac{1}{A}\int\int_A\left(\sqrt{1+f_x^2+f_y^2}-1\right)dxdy
\end{equation}Also the \textit{Sdr} parameter, as the \textit{Sdq} parameter, is dimensionless.
\subsection*{Volume parameters}
\subsubsection{Vv}
The \textit{Vv(mr)} parameter, the Void Volume, is the volume of space bounded by the surface texture from a plane at a height corresponding to a chosen \textit{material ratio} value to the lowest valley. The “\textit{mr}” value may be set to any value from 0\% to 100\%: the \textit{Vv} parameter value will be maximum at $mr=0\%$ and minimum at $mr=100\%$.
\subsubsection{Vvc}
$Vvc(p,q)$, the Core Void Volume, is the volume of space bounded by the texture at heights corresponding to the material ratio values of “$p$” and “$q$”. The default value for “$p$” is 10\% and the default value for “$q$” is 80\%. This parameter is useful to evaluate the empty space between the walls of the surface in a 'central' range of heights.
\subsection*{Distribution of normals}
In this section, the definition of the distribution of normals widely used in the literature\cite{heitz2014understanding} for the evaluation of surfaces is shown. For each measured point of the studied surface, its unit normal vector $\hat{w}$ can be calculated and can be described by spherical coordinates $(\theta,\varphi)$ as
\begin{equation}
\label{eqn:verosre2}
    \hat{w}(\theta,\varphi)=(\sin\theta\cos\varphi,\sin\theta\sin\varphi,\cos\theta)
\end{equation}If the coordinates $\theta$ and $\varphi$ are known for all measured points, then the distribution of these coordinates can be calculated: this distribution is the distribution of normals $D(\theta, \varphi)$ and it is measured in [m$^2\cdot$ st$^{-1}$], where st indicates steradian. Extending the definition to the continuous case, the distribution of the normals is defined in such a way that the portion of area $dS_{\Sigma}(\Omega)$ of the surface $\overrightarrow{S_{\Sigma}}$ indicated in equation (\ref{eqn:1}), whose unit vectors have spherical coordinates between the solid angle $\Omega$ and $\Omega+d\Omega$, can be calculated as
\begin{equation}
\label{eqn:relationomega}
    \begin{split}
        dS(\Omega)&=D(\Omega)d\Omega\\
        &=D(\theta, \varphi)\sin\theta d\theta d \varphi
    \end{split}
\end{equation}The whole area $S_{\Sigma}$ of the Cartesian surface $\overrightarrow{S_{\Sigma}}$ can then be calculated by integrating the relation (\ref{eqn:relationomega}) in the hemisphere of the solid angle $\Omega_{+}$, in which the normal unit vectors have the third Cartesian component always positive. In the literature\cite{heitz2014understanding}, the distribution of normals is normalised so that the area of the projection of the surface $\overrightarrow{S_{\Sigma}}$ on the \textit{x-y} plane, i.e. the sampling area, has an area equal to 1 m$^2$, i.e.
\begin{equation}
    \int\int_{\Omega_+}D(\theta, \varphi)\sin\theta\cos\theta d\theta d \varphi=1\quad\text{[}m^2\text{]}
\end{equation}
\section{Measurement Method}
\label{section:measmethod}
\firstword{T}{his}% Capital letter in first word
section shows the procedure used to calculate the ISO 25178 morphological parameters. The calculation of the parameters is obtained through the software Leica, called \textit{Mountains Map}\cite{Leica}. This software allows to analyze the images obtained both by confocal microscope and SEM. It also allows the application of different morphological filters to the measurements in order to estimate the most representative morphological parameters. \\This procedure is divided into the following steps:
\begin{enumerate}
    \item Obtain the measurement of surface morphology by confocal microscope. This measurement contains the Cartesian coordinates ($x,y,z$) of all points measured by the optical instrument;
    \item In case the area obtained from the measurement does not have a square area, or in case a smaller portion of the area wants to be evaluated, cut out a square portion of the measured area. The obtained portion of square area $l\times l$ will be used for the calculation of morphological parameters;
    \item Apply S filter. A low-pass filter is applied to the square area obtained: this filter has a nesting index equal to $S_1$. In this way, the primary coordinate surface ($x_S,y_S,z_S$) is obtained. This value is a function of the maximum sampling distance and eliminates small-scale components of the surface. The $S_1$ filter nesting index is determined according to the recommendations of the ISO 25178-3:2012;
    \item Remove the form. It can often happen that the sample is not perfectly perpendicular to the optical axis of the confocal microscope, causing errors in the calculation of morphological parameters. Therefore, in this work the shape has always been removed through a geometric plane: this plane is calculated through the method of least squares and the value that this plane takes in the \textit{x} and \textit{y} coordinates of the measured points is subtracted from the \textit{z} coordinates of each point. In this way, the \textit{S-F} surface of coordinates ($x_{SF},y_{SF},z_{SF}$) is obtained.
\end{enumerate}
All morphological parameters calculated and shown in this paper are obtained from the \textit{S-F} surfaces obtained from each measurement. The \textit{S-F} surfaces used have $l=40$ $\mu$m and a nesting index $S_1=0.8$ $\mu$m has been chosen for all samples studied.\\
The confocal microscope used for all measurements is the Leica DCM8\cite{DCM8} and the objective used for the measurements is a 150X. Table \ref{table-1} shows schematically the characteristics of the optical lens used.
\begin{table}[h]
\centering
 \begin{tabular}{ccccc}
  \hline
  \hline
  \thead{Bright field \\ Objectives} & \thead{NA} & \thead{Spatial \\ Sampling [$\mu $m]} & 
                        \thead{Optical \\ Resolution [$\mu$ m]} \\
  \hline
  \hline
  150X & $0.95$ & $0.09$ & $0.14$ \\
  \hline
  \hline
 \end{tabular}
   \caption{General specifications of the 150X objective of Leica DCM8 confocal microscope.} \label{table-1}
\end{table}

\section{Validation of the Software}
\begin{figure}[h]
    \centering
    \includegraphics[scale=.45]{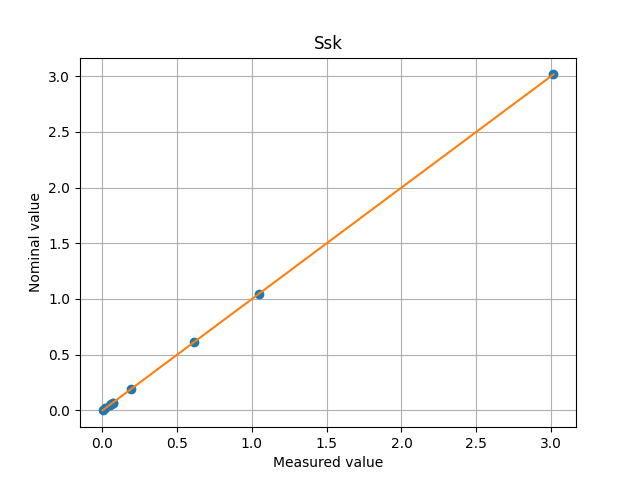}
    \caption{On the \textit{x}-axis, the \textit{Ssk} values measured by the software used were reported and on the \textit{y}-axis the values of the parameter provided by the website of NIST were plotted.}
    \label{fig:ssk}
\end{figure}

\firstword{B}{efore}% Capital letter in first word
proceeding with the measurements through the confocal microscope and the calculation of morphological parameters, the goodness of the software used was verified. A reference dataset extracted from the NIST (National Institute of Standards and Technology) Internet based Surface Algorithm Testing System was used\cite{Nist}. The height parameters of maps SG\_1-1, SG\_1-2, SG\_1-3, SG\_1-4, SG\_2-1, SG\_2-3, SG\_3-1 and SG\_3-2 have been calculated through the Leica Software and compared with the NIST certified data. 
\begin{figure}[h]
    \centering
    \includegraphics[scale=.4]{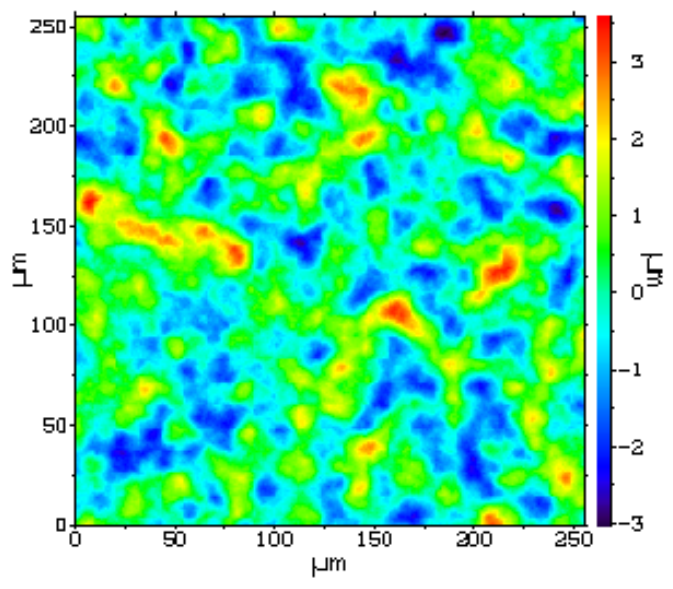}
    \caption{An example of map used to evaluate the goodness of software: in this case, the SG\_1-2 downloaded from the website of NIST is shown.}
    \label{fig:Sg_1-2}
\end{figure}
\\Going into greater detail, for each surface downloaded from the database, the height parameters were calculated\footnote{The parameter \textit{Sq}, \textit{Sp}, \textit{Sv}, \textit{Ssk} and \textit{Sku} were calculated.}. Subsequently, a graph was obtained for each parameter studied, where the values of the parameters obtained and measured with the software used have been reported on the \textit{x}-axis and the values of the parameters provided by the site have been reported on the \textit{y}-axis. In this way, if the software used worked correctly, the expected best fit that correlates the measured parameters with the parameters provided by the NIST is a linear fit of the type $y=mx+q$ whose angular coefficient $m$ is compatible with 1 and the intercept $q$ with 0.\\An explanatory example is reported to make this procedure clearer. For each of the downloaded surfaces\footnote{In this example, unfiltered map was downloaded.}, the \textit{Ssk} parameter was calculated with the software used. The eight measured values of \textit{Ssk} were then obtained and each of these was reported on the \textit{x}-axis of a graph. Similarly, the \textit{Ssk} parameter values corresponding to the relevant maps provided by NIST were plotted on the \textit{y}-axis. The result obtained for the \textit{Ssk} parameter is shown schematically in figure \ref{fig:ssk}.
\\At this point, the intercept and angular coefficient of the best fit shown in the figure \ref{fig:ssk} via an orange line have been calculated. In this case, the values are\footnote{In this case, both parameters obtained are dimensionless.}
\begin{equation}
    \begin{cases}
    m=(1.00009\pm0.00016)\\
    q=(-16\pm18)\cdot10^{-5}
    \end{cases}
\end{equation}As expected both values obtained about the angular coefficient and the intercept are compatible with 1 and 0 respectively: this consideration was decided to be sufficient to consider reliable the values of the \textit{Ssk} parameter calculated by the software used.\\This procedure is repeated for \textit{Sq}, \textit{Sp}, \textit{Sv} and \textit{Sku} parameters using unfiltered surfaces. The results obtained about the best fit parameters are shown schematically in table \ref{table-2}.
For the tables \ref{table-2} and \ref{table-3}, the values of \textit{q} and their errors for the morphological parameters of \textit{Sq}, \textit{Sp} and \textit{Sv} are reported in [$\mu$m].
\begin{table}[h]
\centering
 \begin{tabular}{ccccc}
  \hline
  \hline
  \thead{Parameters} & \thead{$m$} & \thead{$\sigma_m$} & 
                        \thead{$q$} &\thead{$\sigma_q$}\\
  \hline
  \textit{Sq} & $0.9999999$ & $2\cdot 10^{-7}$ & $9\cdot10^{-7}$&$5\cdot10^{-7}$ \\
  \hline
  \textit{Sp} & $1.0000004$ & $5\cdot 10^{-7}$ & $-4\cdot10^{-7}$&$2\cdot10^{-7}$\\
  \hline
  \textit{Sv} & $0.9999998$ & $4\cdot 10^{-7}$ & $3\cdot10^{-7}$&$2\cdot10^{-7}$\\
  \hline
  \textit{Ssk} & $1.00009$ & $16\cdot 10^{-5}$ & $-16\cdot10^{-5}$& $18\cdot10^{-5}$ \\
  \hline
  \textit{Sku} & $1.000007$ & $8\cdot 10^{-6}$ & $-15\cdot10^{-6}$&$6\cdot10^{-6}$\\
  \hline
  \hline
 \end{tabular}
   \caption{Parameters about the best fit obtained for unfiltered maps are shown.} \label{table-2}
\end{table}
The same type of work was similarly obtained for filtered surfaces with nesting index $S_1=0.008$ mm made available from the same website. The results obtained about the best fit parameters are shown schematically in table \ref{table-3}.
\begin{table}[h]
\centering
 \begin{tabular}{ccccc}
  \hline
  \hline
  \thead{Parameters} & \thead{$m$} & \thead{$\sigma_m$} & 
                        \thead{$q$} &\thead{$\sigma_q$}\\
  \hline
  \textit{Sq} & $1.0000002$ & $2\cdot10^{-7}$ & $5\cdot10^{-7}$&$3\cdot10^{-7}$ \\
  \hline
  \textit{Sp} & $0.9999999$ & $5\cdot 10^{-7}$ & $3\cdot10^{-7}$&$3\cdot10^{-7}$\\
  \hline
  \textit{Sv} & $1.0000007$ & $8\cdot 10^{-7}$ & $-3\cdot10^{-7}$&$2\cdot10^{-7}$\\
  \hline
  \textit{Ssk} & $1.00004$ & $9\cdot10^{-5}$ & $-3\cdot10^{-5}$& $5\cdot10^{-5}$ \\
  \hline
  \textit{Sku} & $0.99998$ & $12\cdot10^{-5}$ & $2\cdot10^{-5}$& $4\cdot10^{-5}$ \\
  \hline
  \hline
 \end{tabular}
   \caption{Parameters about the best fit obtained for filtered maps with nesting index $S_1=0.008$ $mm$ are shown.}
   \label{table-3}
\end{table}\\All parameters about the best fit obtained are compatible with the expected value using a confidence interval of $3\sigma$: this observation was considered sufficient condition to consider the software used reliable.
\section{ABS surface measurements}
\label{section-ABSmeas}
\firstword{T}{he}% Capital letter in first word
samples of ABS surfaces studied are six. This set of samples is divided into two groups: 3 samples have been subjected to an etching treatment through chemical baths based on hexavalent chromium, while the other 3 have been subjected to a similar treatment but without using hexavalent chromium. Within each of these two groups, there are three types of samples: the "\textit{Under}" sample, indicating that the immersion time inside the chemical bath is 5 minutes, that is a time relatively low, the "\textit{Standard}" sample, whose for the samples studied the immersion time is 10 minutes and the "\textit{Over}" sample, whose immersion time is relatively high, for immersion times between 15 and 20 minutes. \\
The measurements on ABS samples were made using the microscope and optical objective whose specifications are schematically shown in table \ref{table-1} and the measurement of morphological parameters for each sample was performed as follows. For each sample, 3 different S-F surfaces were obtained: for each of them, the morphological parameters were calculated. The best estimate of the morphological parameter of the sample was the average of the 3 parameters obtained from the respective S-F surface, while the error associated was the semi-dispersion of the 3 measurements. \\In the following, the label "Cr6" indicates samples subjected to hexavalent chromium etching, whereas "Cr6-Free" refers to samples subjected to an alternative industrial etching treatment.\\ Table \ref{table-ABS} summarises schematically the names of the samples used and the time of immersion of the etching process.
\begin{table}[h]
    \centering
\begin{tabular}{|c|c|c|c|c|c|c|c|}
\hline
\hline
\multicolumn{6}{|c|}{\textbf{\Large ABS, Immersion of time [min] }}\\
\hline
\hline
\multicolumn{3}{|c|}{\textbf{"Cr6"}}&\multicolumn{3}{c|}{\textbf{"Cr6-Free"}}\\
\hline
\multicolumn{1}{|c|}{\textbf{\textit{Under}}} &\multicolumn{1}{c|}{\textbf{\textit{Standard}}} &\multicolumn{1}{c|}{\textbf{\textit{Over}}} &\multicolumn{1}{c|}{\textbf{\textit{Under}}} &\multicolumn{1}{c|}{\textbf{\textit{Standard}}} &\multicolumn{1}{c|}{\textbf{\textit{Over}}} \\
\hline
5 & 10&15&5& 10&20  \\
\hline
\end{tabular}
\caption{The names used for the samples and the times of immersion, expressed in minutes, are shown.}
\label{table-ABS}
\end{table}
\subsection*{Comparison of etching process}
In this section, results obtained for ABS surfaces etched through hexavalent chromium and "Cr6-Free" samples are compared.
\subsubsection{Height parameters}
\begin{figure}
    \centering
    \includegraphics[scale=.42]{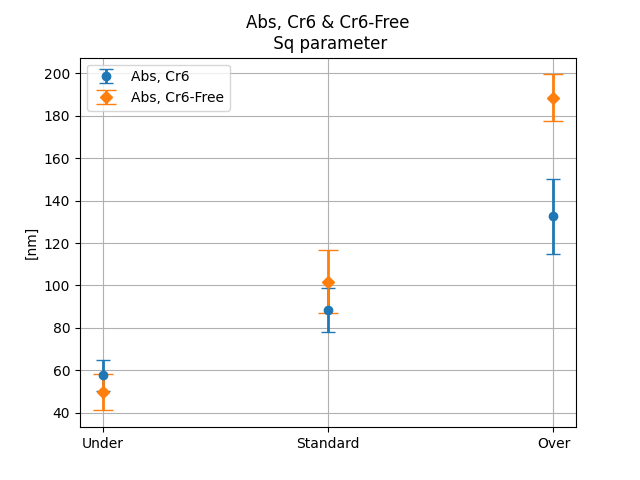}
    \includegraphics[scale=.42]{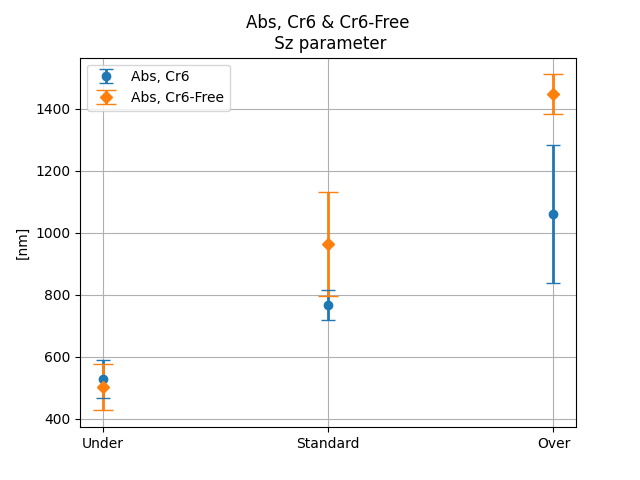}
    \caption{On the top, the \textit{Sq} measurements are shown. On the bottom, the \textit{Sz} measurements are reported.}
    \label{fig:cr6_ABS_height}
\end{figure}

\begin{figure}[h]
    \centering
    \includegraphics[scale=.42]{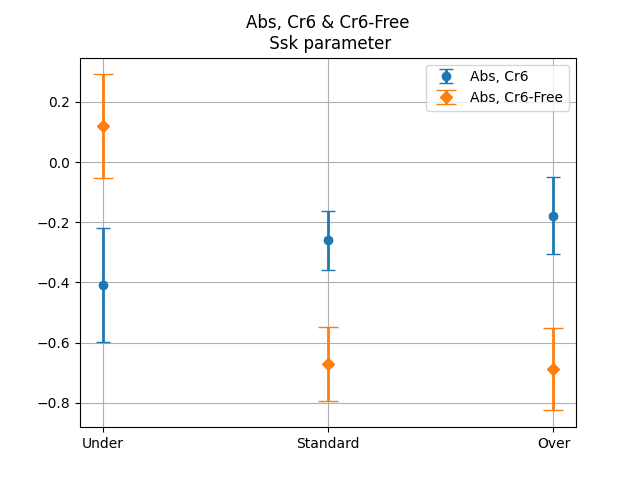}
    \includegraphics[scale=.42]{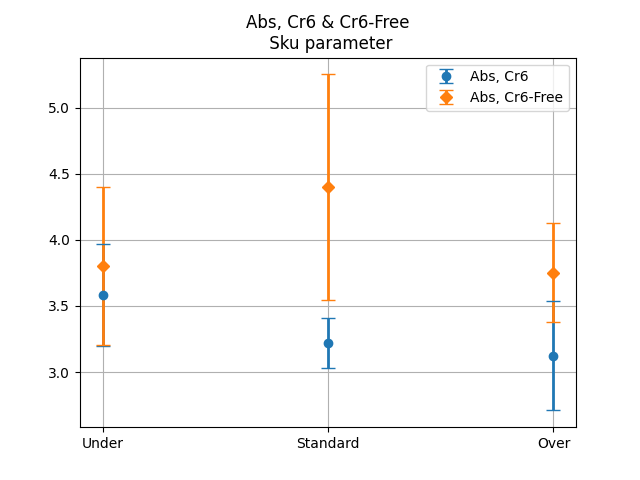}
    \caption{On the top, the \textit{Ssk} measurements are shown. On the bottom, the \textit{Sku} measurements are reported.}
    \label{fig:cr6_ABS_height2}
\end{figure}

The parameters \textit{Sq} and \textit{Sz} are shown in figure \ref{fig:cr6_ABS_height}, while the parameters \textit{Ssk} and \textit{Sku} are reported in figure \ref{fig:cr6_ABS_height2}. \\ As can be observed from figure \ref{fig:cr6_ABS_height}, both \textit{Sq} and \textit{Sz} increase with increasing immersion time: this statement is valid for both the samples. From the same figure, it is possible to observe that the parameters are not compatible for immersion times longer than 10 minutes. This observation may be consistent with what has been observed from the images acquired with the SEM: the difference can be appreciated by observing the images of figure \ref{fig:ABScr6SEM} and \ref{fig:ABSEvolveSEM}.
\\
The \textit{Ssk} and \textit{Sku} parameters are instead shown in figure \ref{fig:cr6_ABS_height2}. As the \textit{Ssk} parameter is defined, it assumes a zero value if the distribution of the heights of the measured points is a normal distribution, it assumes negative values if there is a predominance of valleys and assumes positive values if, on the contrary, there is a predominance of peaks. It is possible to observe from figure \ref{fig:cr6_ABS_height2} that the \textit{Ssk} parameter for the Cr6 ABS surfaces has an almost constant trend and for high immersion times the \textit{Ssk} value approaches zero. For this type of samples, therefore, the distribution of heights becomes slightly more symmetrical with high immersion times, which makes the presence of valleys and peaks more homogeneous. On the contrary, for the Cr6-Free ABS surfaces, the \textit{Ssk} parameter decreases with increasing the time of immersion, showing that in this case the etching process causes larger and deeper valleys in the sample morphology. This observations are consistent comparing the images obtained by SEM: as shown schematically in figure \ref{fig:ABScr6SEM}, the valleys that characterize "\textit{Under}" samples etched with hexavalent chromium gradually become less and less distinguishable due to higher immersion times, on the contrary, for the Cr6-Free samples, large and deeper valleys are more visible for longer times of immersion, as shown in the lower image of figure \ref{fig:ABSEvolveSEM}. 
More difficult considerations can be made for the \textit{Sku} parameter. As shown in figure \ref{fig:cr6_ABS_height2}, the uncertainties associated with measurements of this parameter are relatively high. Both the \textit{Ssk} and \textit{Sku} parameters are defined by higher order powers of the heights of the measured points and this can increase the error of the measurements of these parameters: for this reason, it is often necessary to conduct several measurements for a more precise estimate of these two parameters. In any case, by definition, the \textit{Sku} parameter assumes values greater than 3 for morphologies characterized by very high peaks and particularly deep valleys: in fact, this parameter is often used to characterize surfaces subject to abrasion and corrosion.
\subsubsection{Hybrid parameters}
\begin{figure}[h]
    \centering
    \includegraphics[scale=.42]{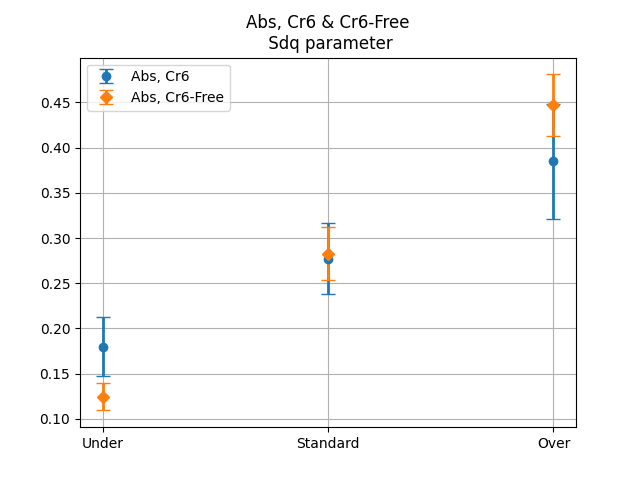}
    \includegraphics[scale=.42]{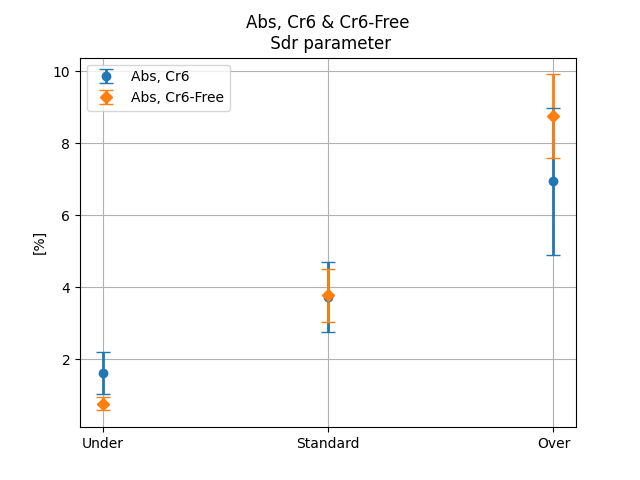}
    \caption{On the top, the \textit{Sdq} measurements are shown. On the bottom, the \textit{Sdr} measurements are reported.}
    \label{fig:cr6_ABS_ibridi}
\end{figure}

The \textit{Sdq} and \textit{Sdr} parameters are shown schematically in figure \ref{fig:cr6_ABS_ibridi}. The hybrid parameters also increase with increasing immersion time of the samples, regardless of the type of etching. In general, the values of the \textit{Sdq} and \textit{Sdr} parameters are high for complex and irregular morphologies and assume lower values for smoother surfaces. Observing figure \ref{fig:cr6_ABS_ibridi}, it is also possible to note that, for immersion times longer than 10 minutes, the \textit{Sdq} and \textit{Sdr} parameters are compatible. 
\subsubsection{Volume parameter}
\begin{figure}[h]
    \centering
    \includegraphics[scale=.42]{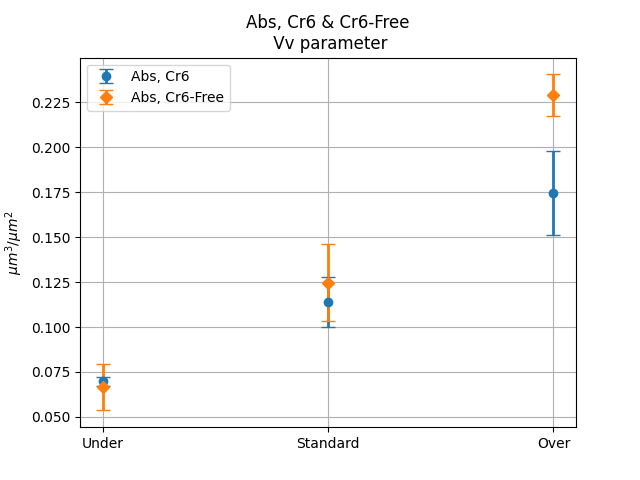}
    \includegraphics[scale=.42]{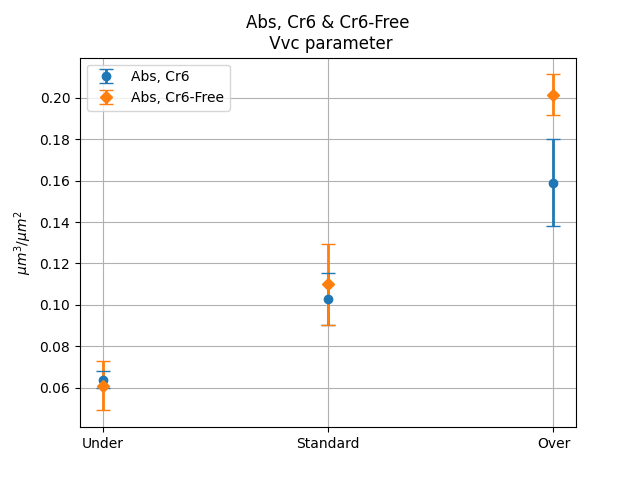}
    \caption{On the top, the \textit{Vv} measurements are shown. On the bottom, the \textit{Vvc} measurements are reported.}
    \label{fig:cr6_ABS_volume}
\end{figure}

The \textit{Vv} and \textit{Vvc} parameters are schematically shown in figure \ref{fig:cr6_ABS_volume}. The \textit{Vv} parameter is a parameter normalised to the measured area and for the measurements shown in this section it was calculated with a material ratio value of $mr=10\%$. Similarly, the \textit{Vvc} parameter was calculated with the values of \textit{p} and \textit{q} set at 10\% and 80\% respectively.\\
As can be seen from figure \ref{fig:cr6_ABS_volume}, these two parameters are compatible for immersion times of less than 10 minutes, regardless of the type of chemical bath: for longer immersion times, the \textit{Vv} and \textit{Vvc} parameters for the Cr6-Free surfaces, however, are larger than the parameters measured on surfaces etched with hexavalent chromium. This statement implies that Cr6-Free treatment causes a larger void volume than the Cr6 treatment in the ABS surfaces.

\section{PC-ABS surface measurements}
\label{section-pcABSmeas}
\firstword{S}{imilarly}% Capital letter in first word
to the ABS surfaces, the PC-ABS samples studied are six. These six samples are divided into two groups: 3 samples were etched using a chemical bath based on hexavalent chromium, while the other 3 were etched using the "Cr6-Free" etching process, i.e. without using hexavalent chromium. Each of the two groups has a sample called "\textit{Under}", which indicates an etching time between 6 and 8 minutes, a sample called "\textit{Standard}" and a sample called "\textit{Over}", which indicates a higher etching time, in this case between 18 and 25 minutes depending on the sample. Table \ref{table-pcABS} schematically shows the names used for the samples studied with the immersion times regarding the etching process.
\begin{table}[h]
    \centering
\begin{tabular}{|c|c|c|c|c|c|c|c|}
\hline
\hline
\multicolumn{6}{|c|}{\textbf{\Large PC-ABS, Immersion of time [min] }}\\
\hline
\hline
\multicolumn{3}{|c|}{\textbf{"Cr6"}}&\multicolumn{3}{c|}{\textbf{"Cr6-Free"}}\\
\hline
\multicolumn{1}{|c|}{\textbf{\textit{Under}}} &\multicolumn{1}{c|}{\textbf{\textit{Standard}}} &\multicolumn{1}{c|}{\textbf{\textit{Over}}} &\multicolumn{1}{c|}{\textbf{\textit{Under}}} &\multicolumn{1}{c|}{\textbf{\textit{Standard}}} &\multicolumn{1}{c|}{\textbf{\textit{Over}}} \\
\hline
6 & 12&18&8&15&25  \\
\hline
\end{tabular}
\caption{The names used for the samples and the times of immersion, expressed in minutes, are shown.}
\label{table-pcABS}
\end{table}\\
The measurements on PC-ABS samples were made using the microscope and optical objective whose specifications are schematically shown in table \ref{table-1} and the morphological parameters for the PC-ABS samples were calculated in a similar way to the ABS samples.
\subsection*{Comparison of etching process}
In this section, the results obtained for PC-ABS surfaces etched through hexavalent chromium and "Cr6-Free" samples are shown.
\subsubsection{Height parameters}
\begin{figure}[H]
    \centering
    \includegraphics[scale=.42]{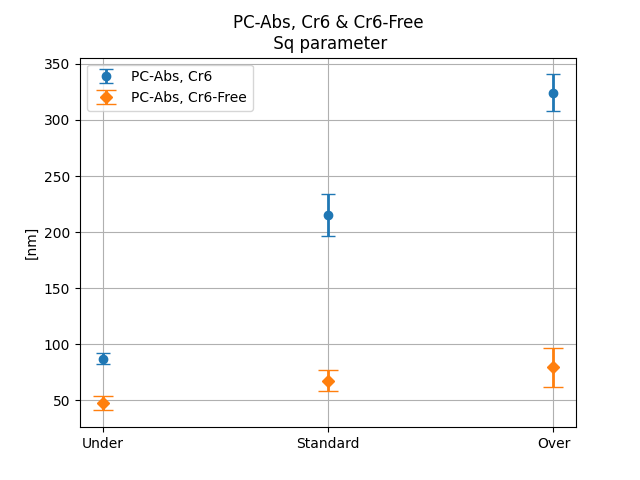}
    \includegraphics[scale=.42]{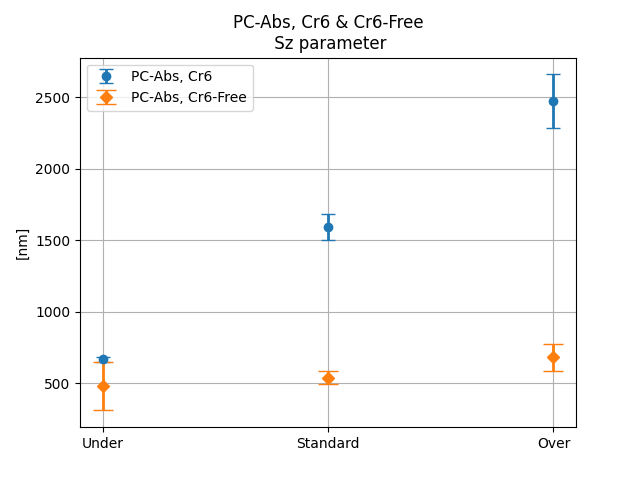}
    \caption{In the image above, the \textit{Sq} parameter measurements are shown. In the bottom image, the measurements of \textit{Sz} are shown.}
    \label{fig:cr6_pcABS_height}
\end{figure}
\begin{figure}[h]
    \centering
    \includegraphics[scale=.42]{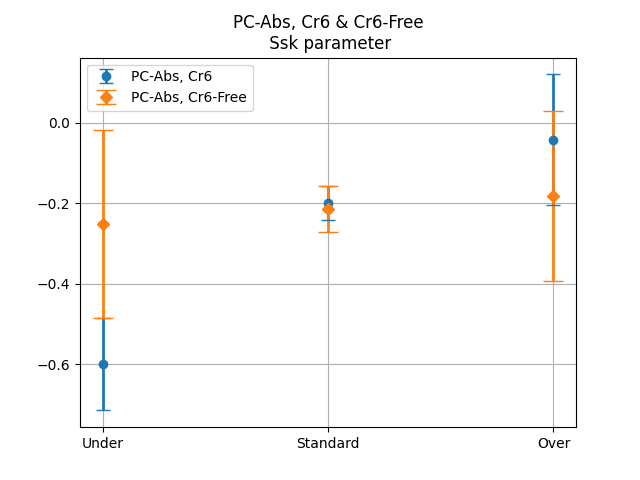}
    \includegraphics[scale=.42]{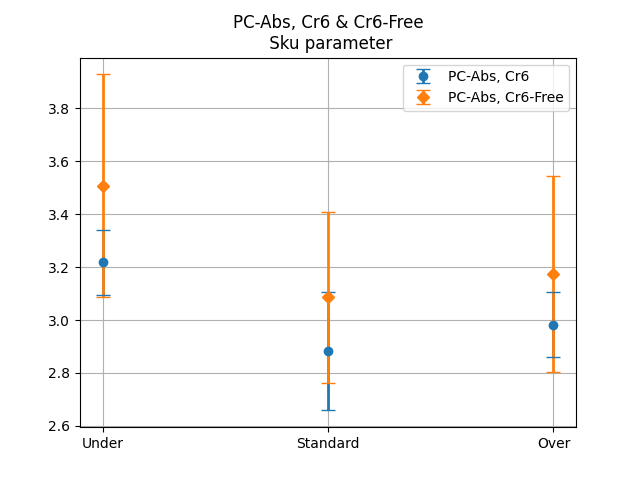}
    \caption{In the image above, the \textit{Ssk} parameter measurements are shown. In the bottom image, the measurements of \textit{Sku} are shown.}
    \label{fig:cr6_pcABS_height2}
\end{figure}

The figure \ref{fig:cr6_pcABS_height} shows the $Sq$ and $Sz$ measurements for different etching times. 
As shown in the same figure, the two etching treatments provide non compatible measurements of $Sq$: the Cr6-Free surfaces are much less sensitive to the etching time than the cr6 surface. Similar results are shown in figure \ref{fig:cr6_pcABS_height} for the $Sz$ parameter. \\
These observations seem to be consistent with the SEM images shown in figures \ref{fig:PCABScr6SEM} and \ref{fig:PCABSEvolveSEM}.
As shown in this images, the effect obtained with a longer immersion time through hexavalent chromium etching is more evident than the "Cr6-Free" etching baths, which more significantly increases the dispersion of the distribution of heights, and consequently the values of \textit{Sq} and \textit{Sz} parameter.\\
A similar study has been done for the \textit{Ssk} and \textit{Sku} parameters, shown in figure \ref{fig:cr6_pcABS_height2}. As shown in the image at the top of figure \ref{fig:cr6_pcABS_height2}, the \textit{Ssk} parameter is always less than zero for both hexavalent chromium-etched and Cr6-Free surfaces.
\\
As for ABS surfaces, the value of \textit{Ssk} parameter for PC-ABS surfaces etched with hexavalent chromium increases with time of immersion, showing that it is more difficult to observe a predominance of surface valleys as the immersion time increases. On the contrary, for the Cr6-Free surfaces, the \textit{Ssk} parameter is slightly lower than zero, but almost constant as a function of immersion time, showing a slight predominance of valleys that remains roughly constant regardless of immersion time. These observations are again confirmed by the SEM images in figures \ref{fig:PCABScr6SEM} and \ref{fig:PCABSEvolveSEM}. Figure \ref{fig:PCABScr6SEM} shows the images of the surfaces etched with hexavalent chromium: as can be seen, the predominance of the valleys disappears as the immersion time increases, as the morphology presents new peaks that are created during the etching process. In figure \ref{fig:PCABSEvolveSEM}, the predominance of valleys remains almost constant as confirmed by the \textit{Ssk} measurements.\\
The \textit{Sku} parameter for both hexavalent chromium-etched and Cr6-Free surfaces is compatible with 3 for etching times longer than 12 minutes ("\textit{Standard}" and "\textit{Over}" samples). These measurements can be interpreted from the definition of this parameter, which assumes values close to 3 when no very high single peaks or very deep single valleys are present along the observed morphology: this observation is also confirmed by the SEM images, through which no particularly very high or very deep singularities are present. As with the parameters measured on the ABS surfaces, the errors associated with the measured values are particularly high and therefore a greater number of measurements may be required to decrease the error associated.
\begin{figure}[h]
    \centering
    \includegraphics[scale=.42]{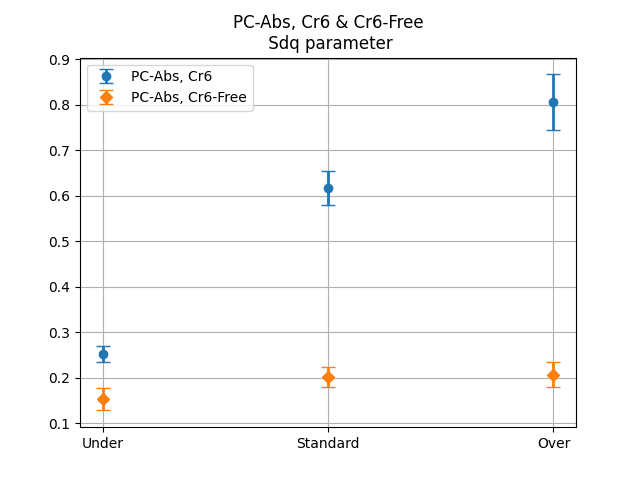}
    \includegraphics[scale=.42]{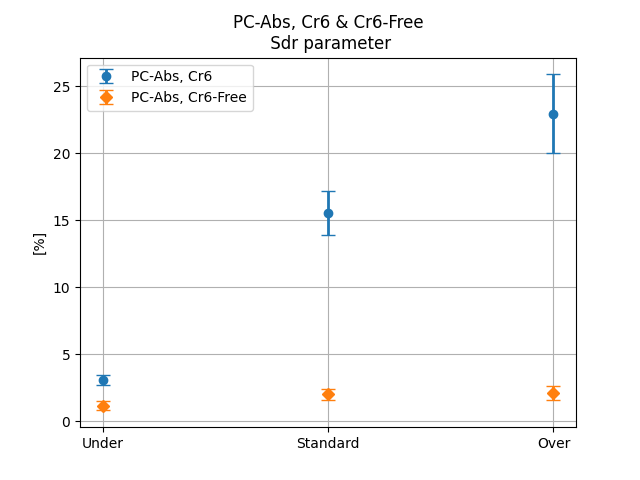}
    \caption{In the image above, the \textit{Sdq} parameter measurements are shown. In the bottom image, the measurements of \textit{Sdr} are shown.}
    \label{fig:cr6_pcABS_ibridi}
\end{figure}
\subsubsection{Hybrid parameters}
The \textit{Sdq} and \textit{Sdr} parameters for PC-ABS surfaces are schematically shown in figure \ref{fig:cr6_pcABS_ibridi}. Similarly to the measurements of the \textit{Sq} and \textit{Sz} parameters, the \textit{Sdq} and \textit{Sdr} parameters also differ significantly as a function of immersion time and the values of \textit{Sdq} and \textit{Sdr} are greater for surfaces etched with hexavalent chromium than for Cr6-Free surfaces. \\
In addition, the \textit{Sdq} value measured on surfaces etched with hexavalent chromium, as well as the \textit{Sdr} value, is not compatible with that measured on the Cr6-Free samples regardless of immersion time, as can be seen from the comparison of the measurements shown in figure \ref{fig:cr6_pcABS_ibridi}.
This statement is again confirmed by the SEM images, where hexavalent chrome-etched surfaces are more irregular than Cr6-Free surfaces. These results can be useful as \textit{Sdq} and \textit{Sdr} are often useful in applications involving surface coatings and adhesion and both parameters find relevance when considering surfaces used with lubricants and other fluids\cite{menezes2008influence}.
\begin{figure}[h]
    \centering
    \includegraphics[scale=.42]{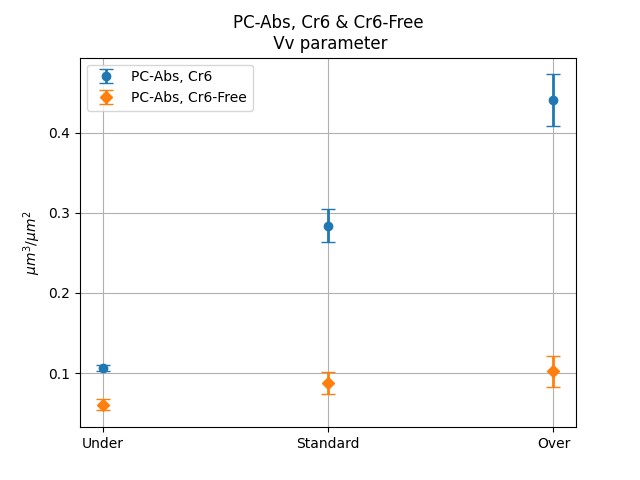}
    \includegraphics[scale=.42]{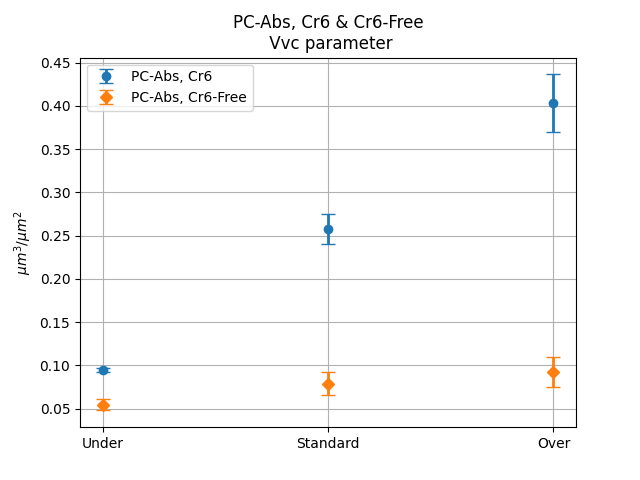}
    \caption{In the image above, the \textit{Vv} parameter measurements are shown. In the bottom image, the measurements of \textit{Vvc} are shown.}
    \label{fig:cr6_pcABS_volume}
\end{figure}
\subsubsection{Volume parameter}
The \textit{Vv} and \textit{Vvc} parameters are shown in figure \ref{fig:cr6_pcABS_volume}. In this section, the \textit{Vv} parameter was calculated with a material ratio equal to $mr = 10\%$. Similarly, also for the PC-ABS measurements, the \textit{Vvc} parameter was calculated with the values of $p=10\%$ and $q=80\%$.
The $Vv$ and $Vvc$ measurements are not compatible for the two etching processes. Furthermore, for surfaces etched with hexavalent chromium, both volume parameters are significantly larger than the parameters measured on the Cr6-Free surfaces and this difference becomes greater as the immersion time increases.\\ These parameters, as the hybrid parameters, are often useful for qualifying surfaces for degree of adhesion by paints or glues.

\section{ABS and PC-ABS comparison}
\begin{figure}[h]
    \centering
    \includegraphics[scale=.42]{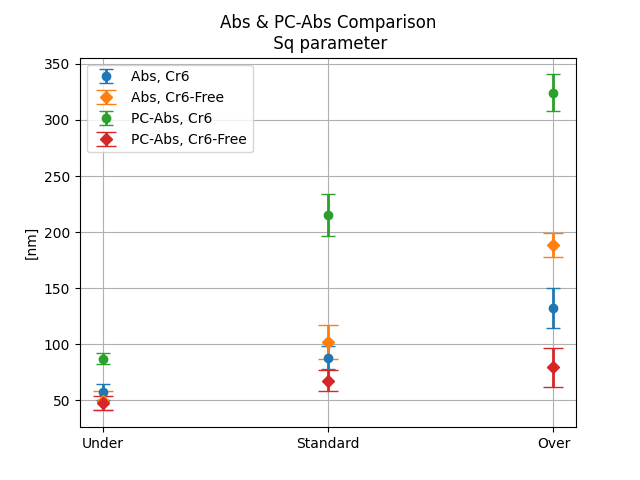}
    \caption{The \textit{Sq} parameter measurements for all samples studied.}
    \label{fig:generalcomparison_SQ}
\end{figure}

\begin{figure}[h]
    \centering
    \includegraphics[scale=.42]{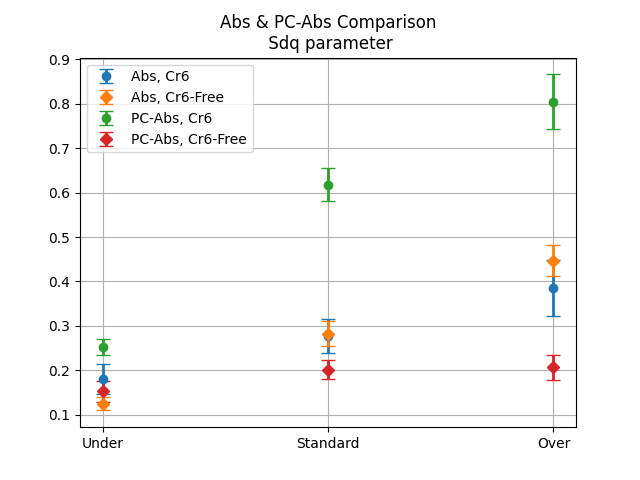}
    \caption{The \textit{Sdq} parameter measurements for all samples studied.}
    \label{fig:generalcomparison_SDQ}
\end{figure}

\begin{figure}[h]
    \centering
    \includegraphics[scale=.42]{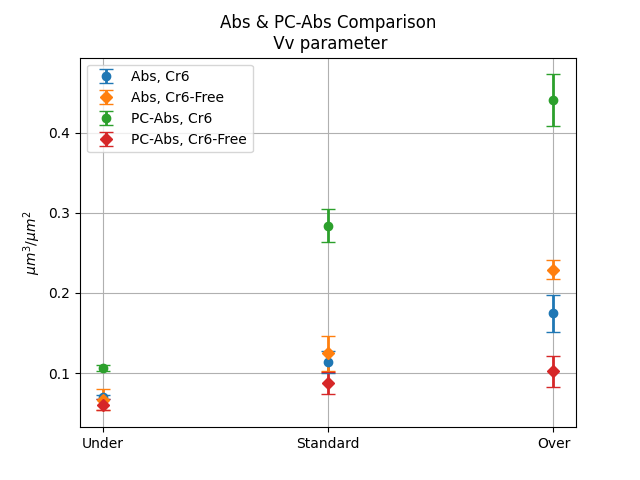}
    \caption{The \textit{Vv} parameter measurements for all samples studied.}
    \label{fig:generalcomparison_Vv}
\end{figure}

The comparison of the measurements of the \textit{Sq}, \textit{Sdq} and \textit{Vv} parameters for all samples studied are shown in figures \ref{fig:generalcomparison_SQ}, \ref{fig:generalcomparison_SDQ} and \ref{fig:generalcomparison_Vv} respectively.\\ The PC-ABS surfaces etched with hexavalent chromium have the highest values of all the parameters, regardless the time of immersion, as shown in figures \ref{fig:generalcomparison_SQ}, \ref{fig:generalcomparison_SDQ} and \ref{fig:generalcomparison_Vv}. It can also be observed that for PC-ABS samples etched with hexavalent chromium, the morphological parameters shown have much larger values than Cr6-Free samples in PC-ABS, despite the fact that the latter were immersed for a longer absolute time, showing a more efficient etching of Cr6 with respect to Cr6-Free treatment. This observation may be able to explain several physical characteristics of such surfaces, such as a better degree of adhesion by applied glues and paints than the other surfaces studied in this work, as these morphological parameters are specifically defined in the ISO 25178 standard to predict such characteristics and the parameters shown are often used for the qualification of surfaces for this type of industrial applications.
The $Sq$, $Sdq$ and $Vv$ parameters relevant to the Cr6-Free treated ABS surfaces are similar to those relevant to the Cr6 treated PC-ABS surfaces. In fact, the values of these three parameters are immediately lower than the parameter values of chrome-etched PC-ABS surfaces. In addition, as can be seen from all three graphs, the \textit{Sq}, \textit{Sdq} and \textit{Vv} parameters increases with the immersion time in the chemical baths: therefore, the Cr6 PC-ABS surface can be reproduced by a long exposure of ABS to Cr6-Free treatment.
\\
As can be seen from figures \ref{fig:generalcomparison_SQ}, \ref{fig:generalcomparison_SDQ} and \ref{fig:generalcomparison_Vv}, for example, the morphological parameters measured for Cr6-Free samples of type "\textit{Over}" in ABS material have values close to the values measured for the PC-ABS surfaces etched with hexavalent chromium but for "Standard" immersion times, i.e. for lower immersion times.
\\
On the other end, the Cr6 ABS surfaces provided similar measurements of these three morphological parameters to the Cr6-Free ABS surfaces, as for immersion times lower than 10 minutes, the parameters are compatible regardless of the type of etching process.\\
In the end, the use of hexavalent chromium is almost indispensable for etching processes for PC-ABS surfaces. In fact, observing the values of the morphological parameters measured for the Cr6-Free PC-ABS surfaces, it is possible to observe how the etching process does not have a great impact on the morphological characteristics of the surface: as shown in the SEM images, the Cr6-Free surface does not change radically from a morphological point of view, regardless of the immersion time of the etching process.

\section{Correlation of parameters}
\begin{figure}[h]
    \centering
    \includegraphics[scale=.5]{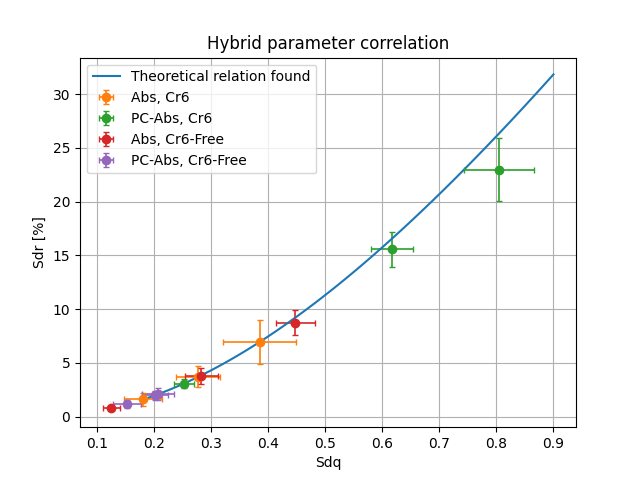}
    \caption{The \textit{Sdr} values as a function of \textit{Sdq} for the studied surfaces in ABS and PC-ABS material. The continuous line is the curve given by equation (\ref{eqn:sdqsdrrrr}).}
    \label{fig:SdrVSSdq}
\end{figure}

\begin{figure}[h]
    \centering
    \includegraphics[scale=.44]{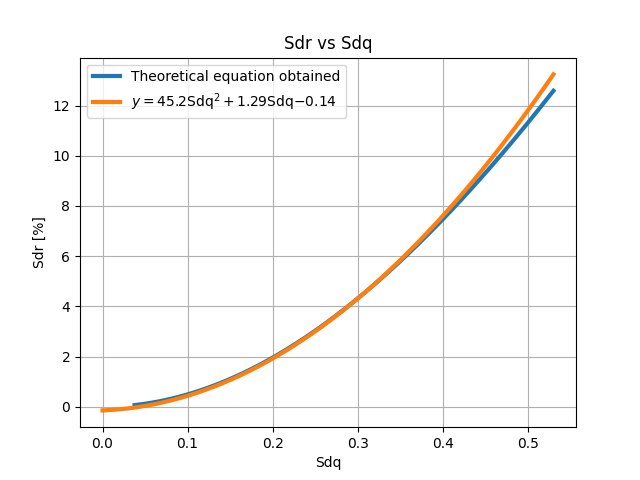}
    \caption{A comparison between the best fit found in the cited article and the theoretical relationship proposed in equation (\ref{eqn:sdqsdrrrr}).}
    \label{fig:SdrVSSdq_articolo}
\end{figure}

\firstword{I}{n} the literature, there are many studies on the possible correlation between different morphological parameters of the ISO 25178 standard. This works show that some parameters of this standard often have a correlation for certain types of surfaces\cite{czifra2020sdq, franco20153d, qi2015correlational}: all studies propose empirical relationships and best fits found through direct measurements on the studied samples, without however proposing theoretical relationships able to explain in detail the experimental evidence. The most studied case is the possible correlation between the \textit{Sdq} and \textit{Sdr} hybrid parameters: observing the definitions of these parameters, it is possible to guess a possible correlation between the two, even if the existence of a real analytical correlation between the two definitions is not obvious. In Appendix \ref{appendix:A} of this paper, a theoretical relation is proposed and it clearly shows how \textit{Sdr} can be written as a function of \textit{Sdq}, assuming that the morphology of the studied surface can be described by the isotropic Beckmann distribution of normals. \\
According to this assumption, it is possible to write the \textit{Sdr} parameter as a function of \textit{Sdq} through the relation
\begin{equation}
\label{eqn:sdqsdrrrr}
    \textit{Sdr} = \frac{\sqrt{\pi}}{2}\cdot\textit{Sdq}\cdot\exp\left\{\frac{1}{\textit{Sdq}^2}\right\}\cdot\text{erfc}\left\{\frac{1}{\textit{Sdq}}\right\}
\end{equation}with $\text{erfc}(x)=1-\text{erf}(x)$, where $\text{erf}(x)$ is the error function.\\ 
In this paper, it was not studied if the Beckmann distribution adequately fits the distribution of normals observed for the studied surfaces, but it is shown that the relationship between the two parameters is evident even for etched surfaces in ABS and PC-ABS materials.\\ In figure \ref{fig:SdrVSSdq}, the values of the measured parameters of \textit{Sdq} are shown on the \textit{x}-axis and the values of \textit{Sdr} on the \textit{y}-axis. In more detail, the value of \textit{Sdq} and the value of \textit{Sdr} were measured for each surface studied. \\
The figure \ref{fig:SdrVSSdq} therefore shows, without reporting a detailed statistical study in this respect, a fit of the experimental data through the theoretical relationship found, both for ABS and PC-ABS surfaces, independently of the etching process used.
Another confirmation of the theoretical relationship found can be obtained for example in \cite{czifra2020sdq}. In this cited work, the values of $Sdr$ and $Sdq$ were measured for different surfaces and the best fit correlating these two parameters was calculated: this best fit is therefore a purely empirical relationship, in contrast to the theoretical relationship proposed and shown in equation (\ref{eqn:sdqsdrrrr}), which was obtained through theoretical evaluations. \\
In figure \ref{fig:SdrVSSdq_articolo}, a comparison between the theoretical relation found in this work reported in equation (\ref{eqn:sdqsdrrrr}) and shown in blue, and the best fit found and proposed by article \cite{czifra2020sdq} that empirically correlates \textit{Sdr} and \textit{Sdq} parameter is shown in orange: also in this case, without reporting a particular statistical study to compare the two relationships, it can be observed that the relationship obtained here is similar to that proposed by the work cited.

\section{Conclusions}
In this paper, it was shown how the morphological parameters defined in the ISO 25178 standard can qualify ABS and PC-ABS surfaces depending on the etching process. Furthermore, the morphological differences that can be achieved depending on the type of material and the type of etching process used were described in detail. \\In the end of work, the correlation between the \textit{Sdq} and \textit{Sdr} parameters for this type of etched surface was studied and a theoretical relationship between the \textit{Sdq} parameter with the \textit{Sdr} parameter was proposed, comparing this result with the measurements obtained on this type of surface and the results obtained in the literature.

% Include appendices (if necessary)
\appendix
\section{Appendices}
\label{appendix:A}
In this appendix, a mathematical demonstration of the relationship shown in equation x is provided, linking \textit{Sdr} as a function of \textit{Sdq} with the assumption that the morphology of the surface studied is described by the isotropic Beckmann distribution of normals in the case of continuous.\\
Let $f:\mathbb{R}^2\to\mathbb{R}$ of class $C^1$ defined in the open set $\Omega$ and let $\overrightarrow{S_{\Sigma}}$ be the Cartesian surface defined as
\begin{equation}
    \overrightarrow{S_{\Sigma}}=(x,y,f(x,y))
\end{equation}In this way, each measured point $P_m$ of the scanned surface can be indicated by the surface $\overrightarrow{S_{\Sigma}}$ and for each of these points its normal versor 
\begin{equation}
\label{eqn:verosre}
    \hat{w}_m(\theta_m,\varphi_m)=(\sin\theta_m\cos\varphi_m,\sin\theta_m\sin\varphi_m,\cos\theta_m)
\end{equation}
can be calculated. Let $S_{\Sigma}$ be the area of the parametric surface $\overrightarrow{S_{\Sigma}}$ and let $S$ be the area of the projection of the surface $\overrightarrow{S_{\Sigma}}$ on the \textit{x-y} plane.\\
In the literature, the distribution of normals is defined in such a way that 
\begin{equation}
    a\int_0^{2\pi}\int_0^{\frac{\pi}{2}} D(\theta,\varphi)\sin\theta d\theta d\varphi=\frac{S_{\Sigma}}{S}
\end{equation}where $a=1$ [m$^{-2}$] is a constant that serves to bring the equation back from a dimensional point of view and it will not be reported from this point.\\Using the definition of the area of a Cartesian surface, it can be shown that the definition of the \textit{Sdr} parameter is equivalent to the following definition
\begin{equation}
    Sdr=\frac{S_{\Sigma}}{S}-1
\end{equation}and therefore the \textit{Sdr} parameter can be calculated as
\begin{equation}
\label{eqn:SdrDistribuzioneeee}
    Sdr=\int_0^{2\pi}\int_0^{\frac{\pi}{2}} D(\theta,\varphi)\sin\theta d\theta d\varphi-1
\end{equation}Additionally, the distribution of normals in literature is normalised so that\cite{heitz2014understanding}
\begin{equation}
\label{eqn:normaliz}
    \int_0^{2\pi}\int_0^{\frac{\pi}{2}} D(\theta,\varphi)\sin\theta\cos\theta d\theta d\varphi=1
\end{equation}On the other hand, the \textit{Sdq} parameter is defined as the square root of the mean value of the square modulus of the gradient, i.e. it is defined as
\begin{equation}
\label{eqn:sdqDistr}
    Sdq=\sqrt{\text{<}\tan^2\theta_m\text{>}}
\end{equation}where $\theta_m$ is the same coordinate shown in equation (\ref{eqn:verosre}). In this way, using the equations (\ref{eqn:normaliz}) and (\ref{eqn:sdqDistr}), it is possible to calculate the \textit{Sdq} parameter through the distribution of normals as
\begin{equation}
    \label{eqn:SdqDistribuzionee}
    Sdq=\sqrt{\int_0^{2\pi}\int_0^{\frac{\pi}{2}} D(\theta,\varphi)\sin\theta\cos\theta\tan^2\theta d\theta d\varphi}
\end{equation}
If the morphology of the studied surface is described through the isotropic Beckmann distribution defined in literature as\footnote{The $\sigma$ quantity is a parameter that defines the morphology of the surface.}
\begin{equation}
    D(\theta,\varphi)=\frac{\exp\left\{-\frac{\tan^2\theta}{\sigma^2}\right\}}{\pi\sigma^2\cos^4\theta}
\end{equation}then the integrals in equation (\ref{eqn:SdrDistribuzioneeee}) and (\ref{eqn:SdqDistribuzionee}) can be calculated, allowing \textit{Sdq} to be obtained as
\begin{equation}
    Sdq=\sigma
\end{equation}and \textit{Sdr} as
\begin{equation}
\label{eqn:sdqsdrrrr_app}
    \textit{Sdr} = \frac{\sqrt{\pi}}{2}\cdot\sigma\cdot\exp\left\{\frac{1}{\sigma^2}\right\}\cdot\text{erfc}\left\{\frac{1}{\sigma}\right\}
\end{equation}In this way, for surfaces described by the isotropic Beckmann distribution, it is possible to write \textit{Sdr} as a function of \textit{Sdq} as
\begin{equation}
\label{eqn:sdqsdrrrr_app2}
    \textit{Sdr} = \frac{\sqrt{\pi}}{2}\cdot\textit{Sdq}\cdot\exp\left\{\frac{1}{\textit{Sdq}^2}\right\}\cdot\text{erfc}\left\{\frac{1}{\textit{Sdq}}\right\}
\end{equation}
The equation found is consistent from a dimensional point of view.

\insertbibliography{References}
\newpage
.
\newpage
\section{SEM images}
\label{sectio_IMMAGINI_SEM}
In this section, the images acquired by SEM are reported.\\
\\
\\

\begin{figure}[h]
    \centering
    \includegraphics[scale=.26]{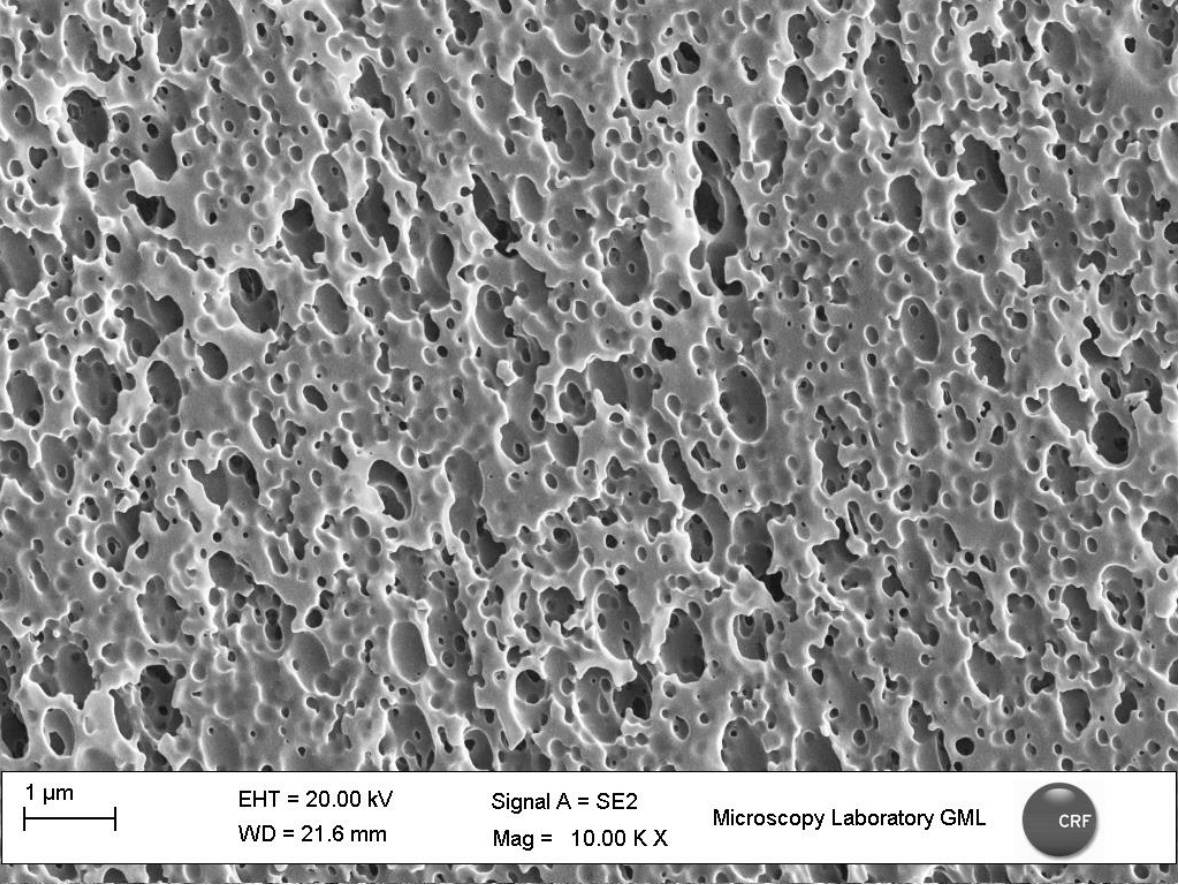}
    \includegraphics[scale=.26]{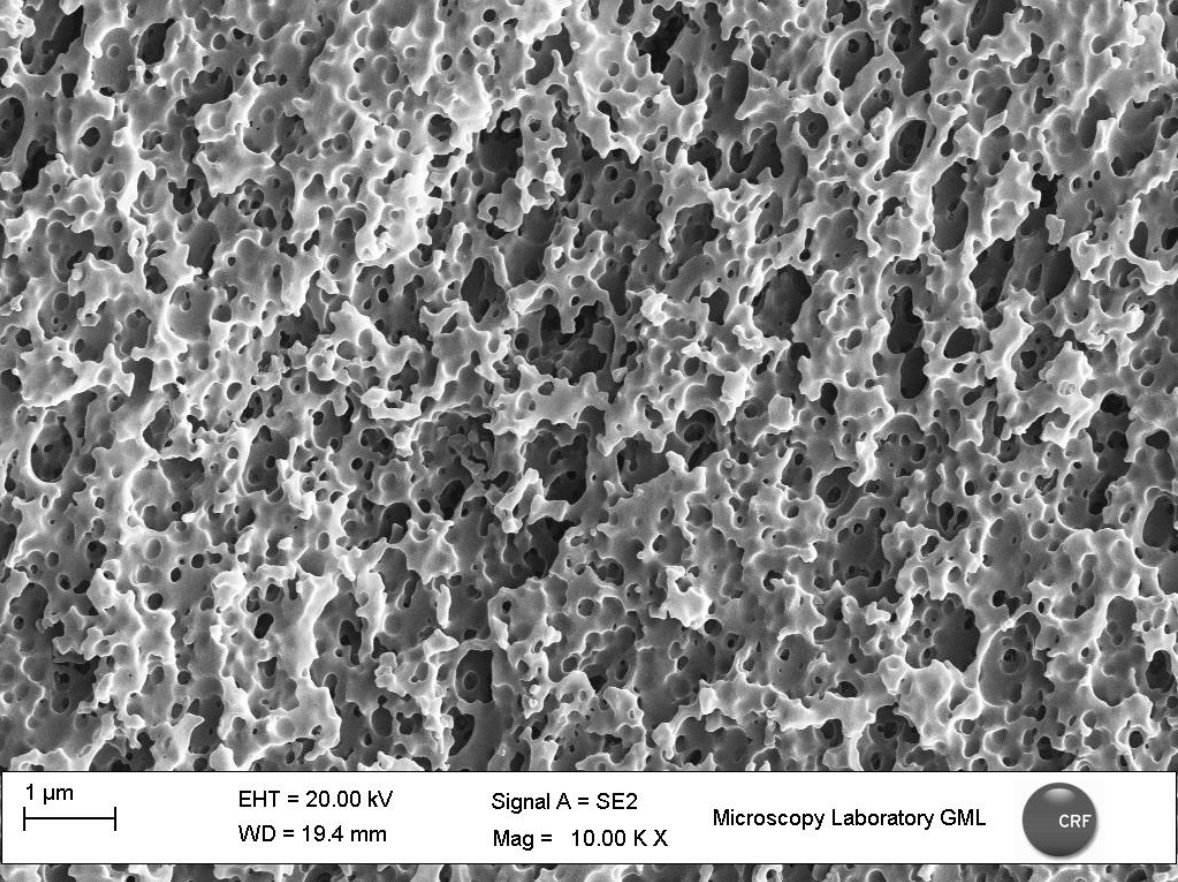}
    \includegraphics[scale=.26]{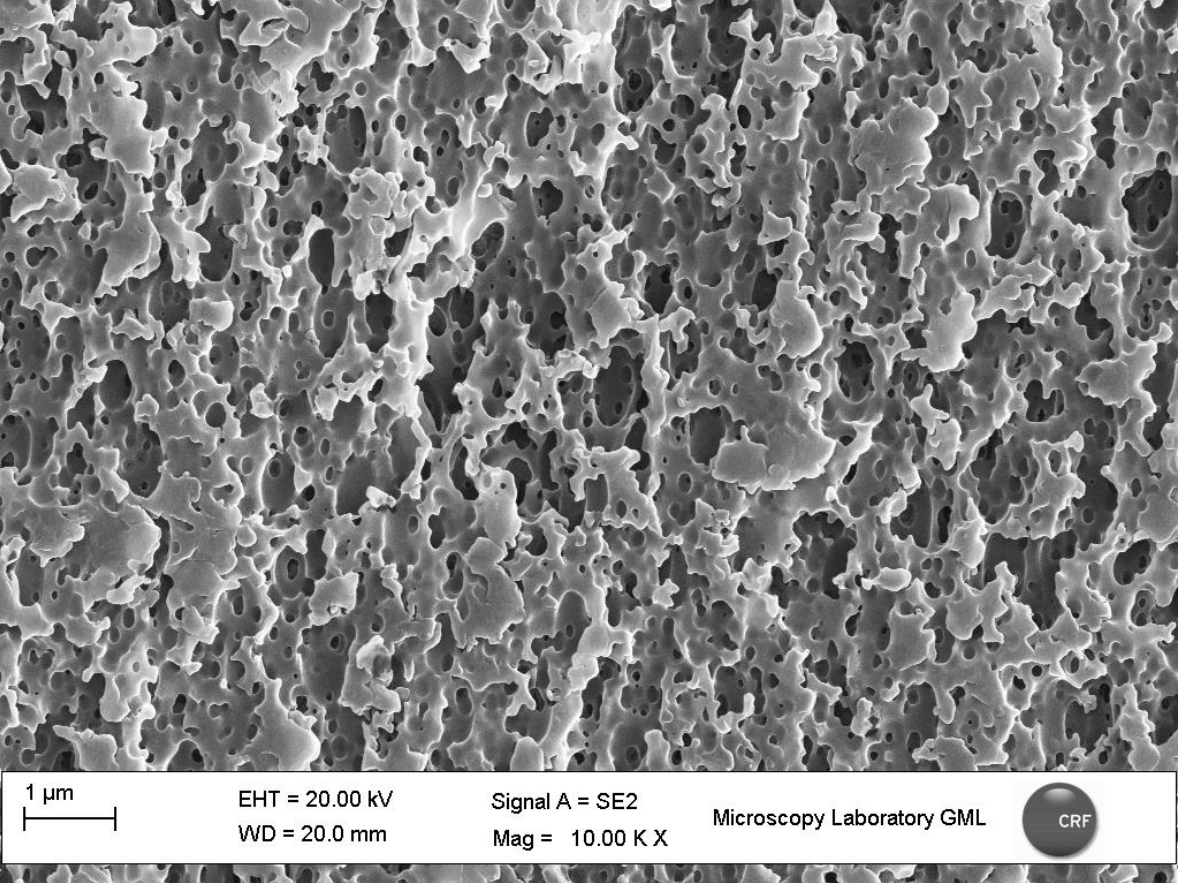}
    \caption{The ABS surfaces etched with hexavalent chromium are shown. The top surface is the "Under" surface (low immersion time), in the middle the "Standard" surface and at the bottom the "Over" surface (high immersion time).}
    \label{fig:ABScr6SEM}
\end{figure}
\begin{figure}[h]
    \centering
    \includegraphics[scale=.26]{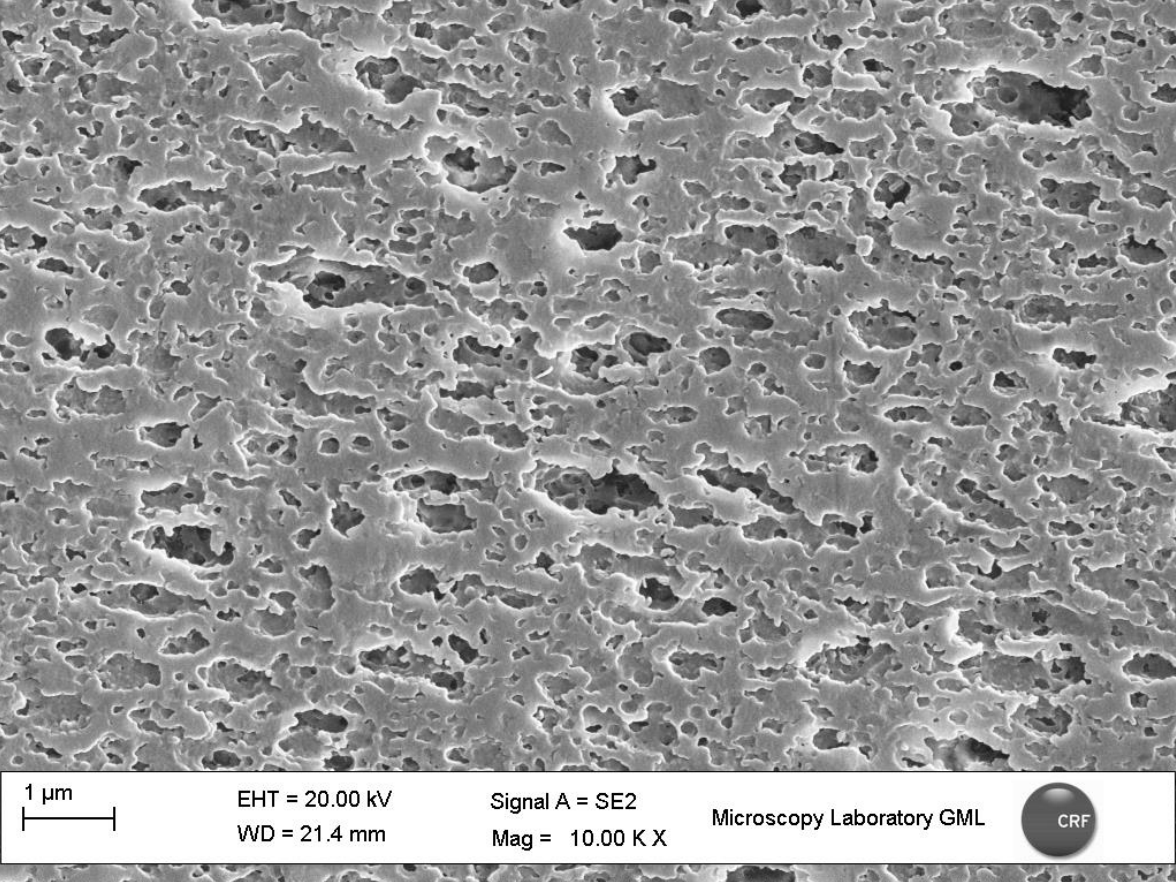}
    \includegraphics[scale=.26]{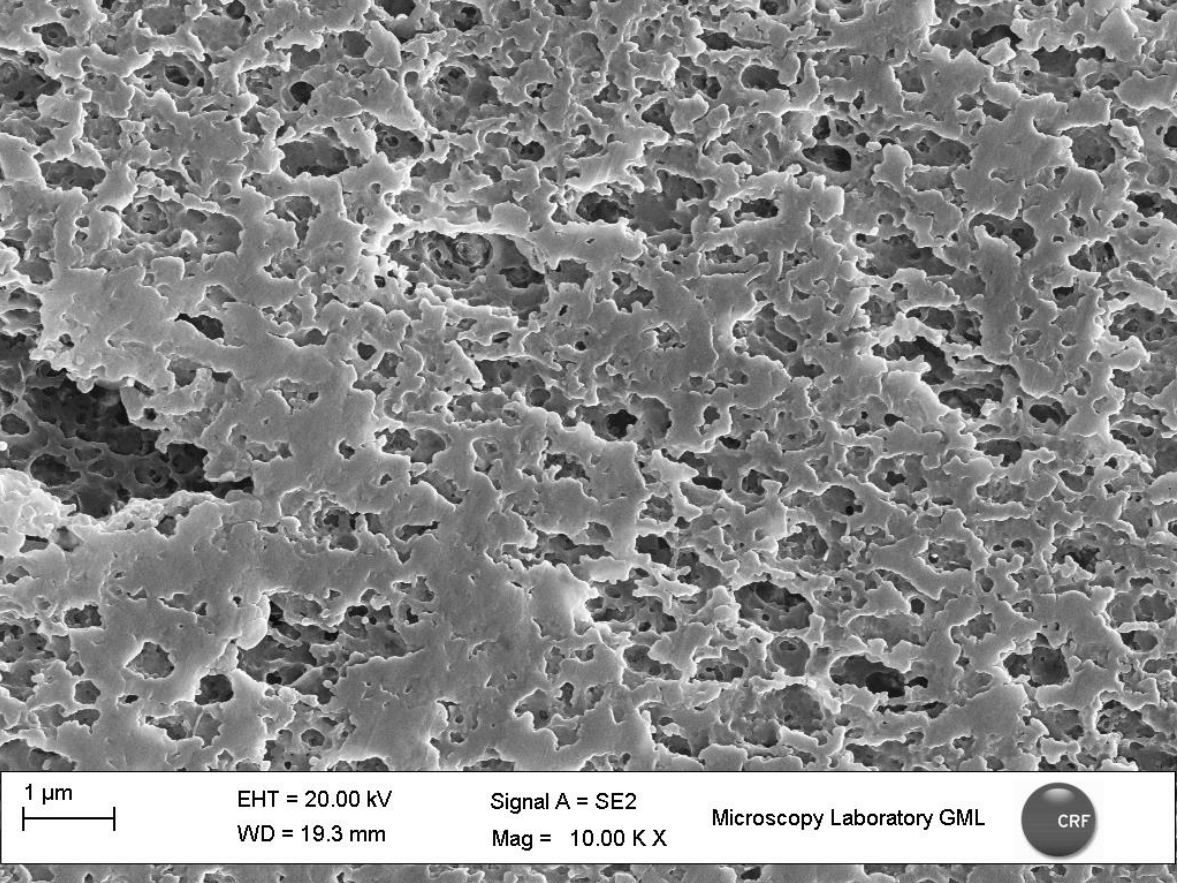}
    \includegraphics[scale=.26]{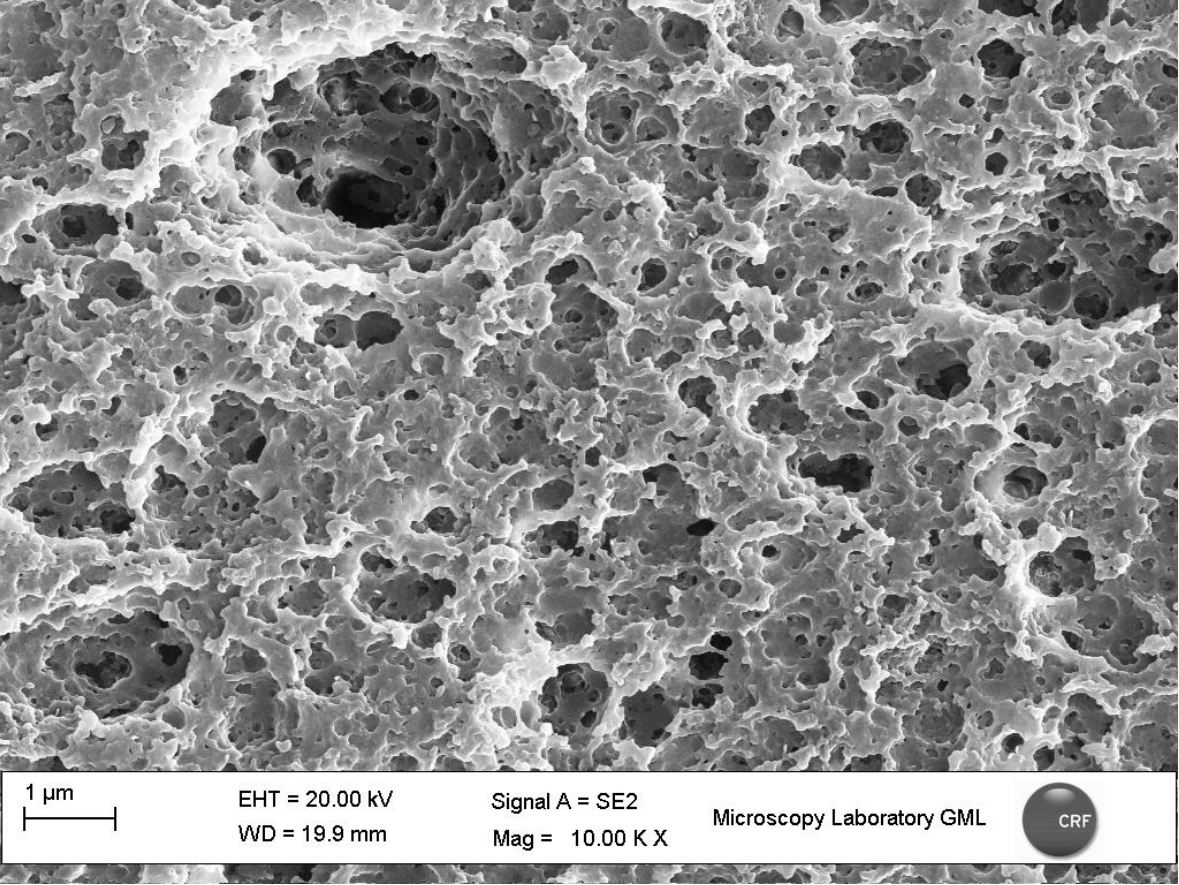}
    \caption{The ABS surfaces etched without hexavalent chromium ("Cr6-Free" surfaces) are shown. The top surface is the "Under" surface (low immersion time), in the middle the "Standard" surface and at the bottom the "Over" surface (high immersion time).}
    \label{fig:ABSEvolveSEM}
\end{figure}
\begin{figure}[h]
    \centering
    \includegraphics[scale=.4]{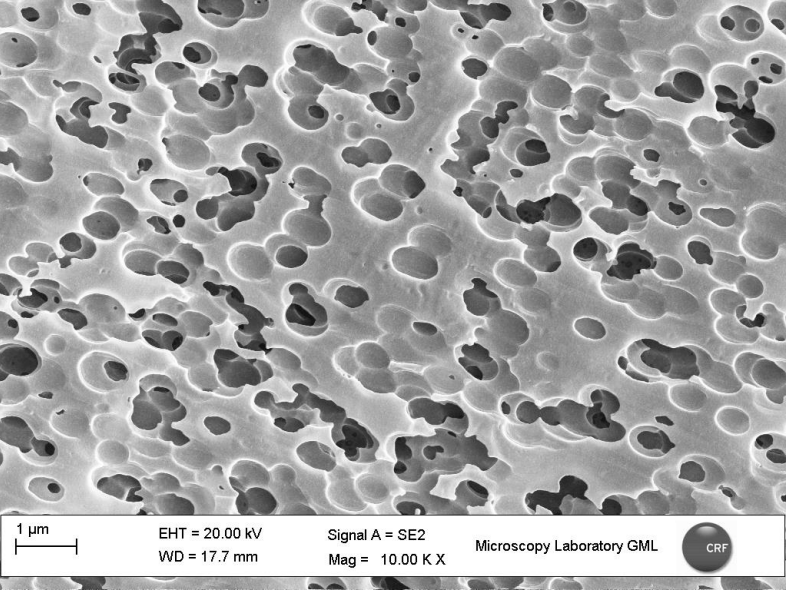}
    \includegraphics[scale=.4]{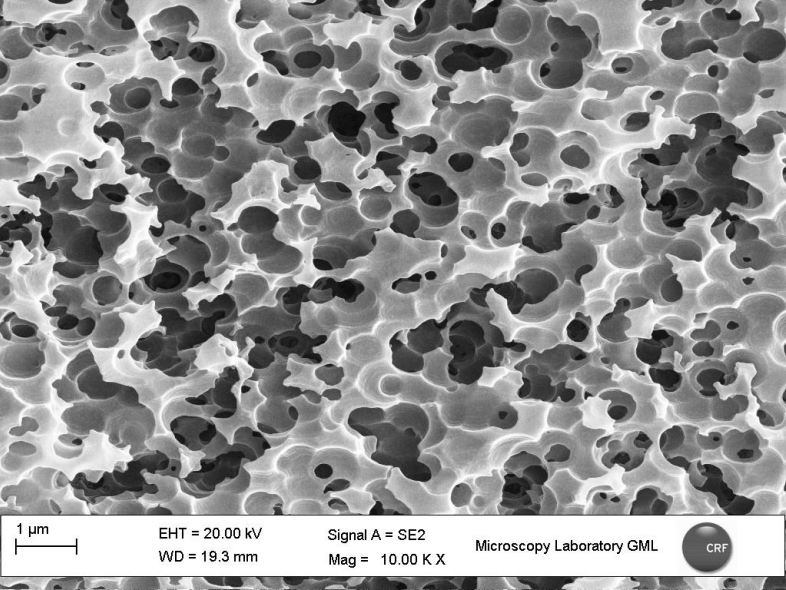}
    \includegraphics[scale=.4]{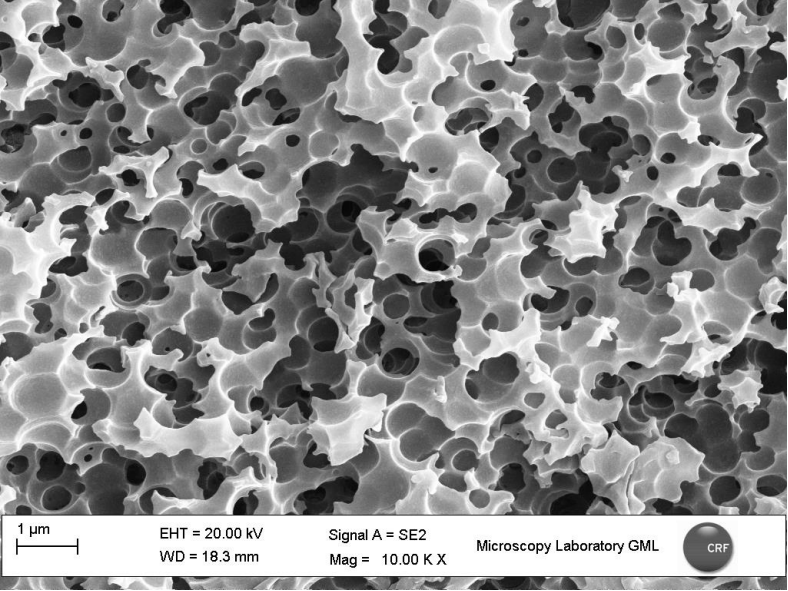}
    \caption{The PC-ABS surfaces etched with hexavalent chromium are shown. The top surface is the "Under" surface (low immersion time), in the middle the "Standard" surface and at the bottom the "Over" surface (high immersion time).}
    \label{fig:PCABScr6SEM}
\end{figure}
\begin{figure}[h]
    \centering
    \includegraphics[scale=.4]{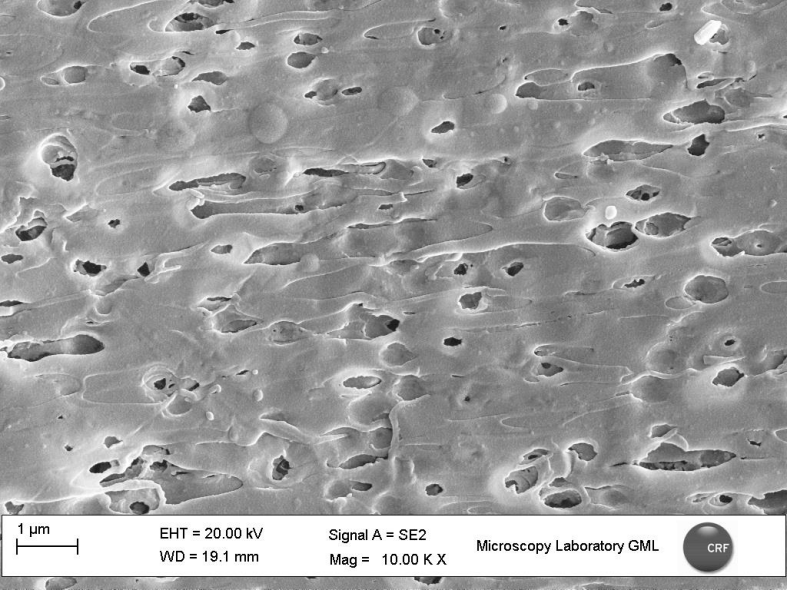}
    \includegraphics[scale=.4]{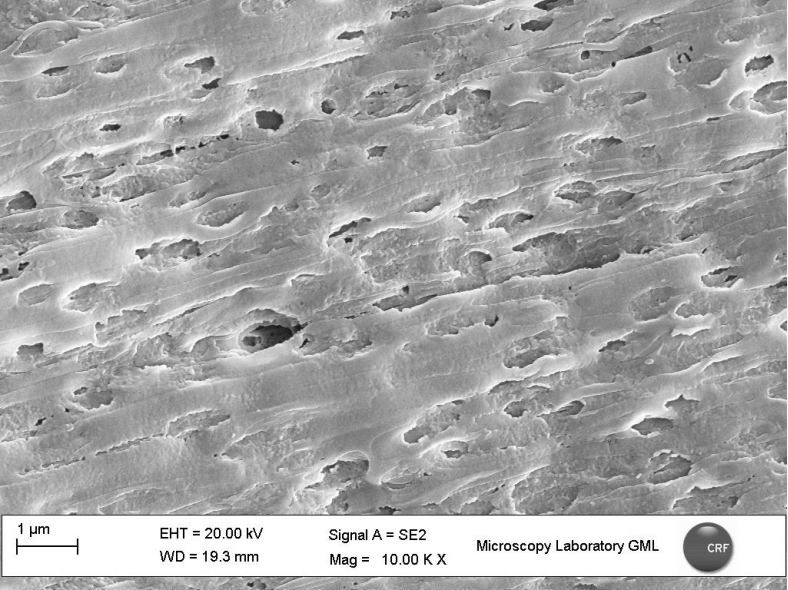}
    \includegraphics[scale=.4]{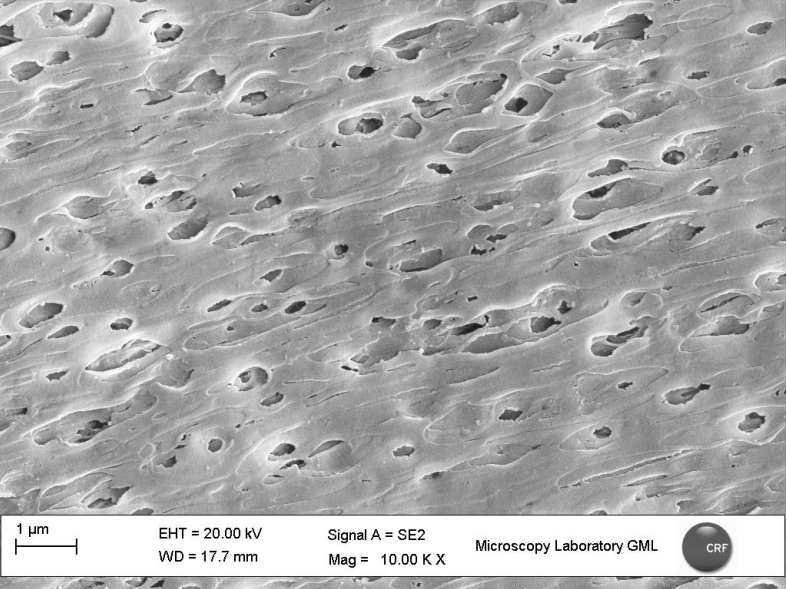}
    \caption{The PC-ABS surfaces etched without hexavalent chromium ("Cr6-Free" surfaces) are shown. The top surface is the "Under" surface (low immersion time), in the middle the "Standard" surface and at the bottom the "Over" surface (high immersion time).}
    \label{fig:PCABSEvolveSEM}
\end{figure}

\end{document}